\documentclass[nobibnotes,nofootinbib,superscriptaddress,showkeys]{revtex4}
\usepackage{amsmath,amssymb,bm}
\usepackage{graphicx}

\newcommand{\beeq}{\begin{eqnarray}}
\newcommand{\eeeq}{\end{eqnarray}}
\newcommand{\be}{\begin{equation}}
\newcommand{\ee}{\end{equation}}
\newcommand{\bea}{\begin{array}}
\newcommand{\eea}{\end{array}}

\newcommand{\eq}{&=&}

\def\dbar{\overline{d}}
\def\ubar{\overline{u}}

\def\funp{{I\!\!P}}
\def\pom{{I\!\!P}}

\begin{document}
\title{\bf Electroweak vector boson production at the LHC as a probe of mechanisms of diffraction}

\author{Krzysztof Golec-Biernat}
\affiliation{Institute of Nuclear Physics Polish Academy of Sciences, Krak\'ow, Poland}
\affiliation{Institute of Physics, University of Rzesz\'ow, Rzesz\'ow, Poland}
\author{Christophe Royon}
\affiliation{CEA Saclay, Irfu/SPP, 91191 Gif/Yvette Cedex, France}
\author{ Laurent Schoeffel}
\affiliation{CEA Saclay, Irfu/SPP, 91191 Gif/Yvette Cedex, France}
\author{ Rafal Staszewski}
\affiliation{Institute of Nuclear Physics Polish Academy of Sciences, Krak\'ow, Poland}

\begin{abstract}
We show that the  double diffractive electroweak vector boson production in the $pp$ collisions  at the LHC is an ideal probe of QCD based mechanisms of diffraction. 
Assuming the resolved Pomeron  model with flavor symmetric parton distributions, the $W$ production
asymmetry in rapidity equals exactly zero. In other approaches, like  the soft color interaction model, in which  soft gluon exchanges are responsible for diffraction,  the asymmetry is non-zero and equal to that in
the inclusive $W$ production. In the same way, the ratio of the $W$ to $Z$ boson production is independent of rapidity in the models with resolved Pomeron  in contrast to the predictions of the soft color interaction model.

\end{abstract}

\maketitle

\section{Introduction}
\label{chapter:1}

Measurements of the charge asymmetry of leptons originating from the decay
of singly produced $W^\pm$ bosons at $pp$, $p\bar{p}$ and $ep$ colliders provides important information
about the proton structure as described by parton distribution functions (PDFs)
\cite{Abe:1994rj,Abazov:2008qv,Aaltonen:2009ta,atlas,cms,Catani:2009sm}.  
In particular, a direct access to these distributions is provided by the measurement of the $W^\pm$ bosons production asymmetry  in rapidity. This quantity reflects the fact that at given rapidity the two charged vector bosons are produced by quarks of different flavors.

In the double diffractive exchange,  the two colliding hadrons remain intact 
creating two gaps in rapidity where no hadronic activity is expected. 
Consequently,
the production of electroweak vector bosons in such processes is done in 
association with the two gaps in rapidity. The production of electroweak 
vector bosons is a QCD process with a hard scale, given by the boson mass, 
thus it can be described perturbatively. However, the QCD based understanding 
of the rapidity gap formation remains a challenge \cite{GolecBiernat:2009pj,general,kupco}.
We should mention at this point that in the 
measurements of diffractive events at the LHC the key element is to tag 
the forward scattered incoming protons.
Because of the pile up of events in each bunch crossing (at high luminosities
up to 35 pile up events per bunch crossing occur), the rapidity gap tagging 
is no longer possible, and the only possibility to detect double diffractive
events is by tagging the intact protons in the final state.

In the resolved Pomeron model interpretation \cite{RPE},  the scattered protons stay 
intact and rapidity gaps are created due to  exchange of two Pomerons
with a partonic structure. Thus,  the electroweak bosons are diffractively 
produced from the annihilation of two quarks  coming from  each of the 
two Pomerons. In other models, like the soft color interaction model \cite{SCI}, 
the diffractive gaps are produced due to soft gluon exchanges which neutralize 
color in the rapidity space between
two outgoing protons and the diffractive system. In this type of models, the 
hard process is independent of the gap formation mechanisms and is the same 
as in the non-diffractive events \cite{GolecBiernat:2009pj,general}.

In order to discriminate between these two essentially different mechanisms 
of diffraction \cite{RPE, SCI} we propose to study
the $W^\pm$  rapidity asymmetry for double diffractive exchanges. In the 
Pomeron exchange models,
the quark content of the Pomeron is flavor symmetric in order to account 
for the vacuum quantum number exchange
between the scattered protons and the diffractive system. Therefore, the 
$W$ production asymmetry is exactly equal to zero. This stays in the contrast 
to the result from the soft color interaction models in which the $W$ 
asymmetry is the same as in the inclusive case,  determined by the $u$ and 
$d$ quark content of the proton.
In the same spirit, the $W$ to $Z$ production rate can be studied, showing 
independence of rapidity in the Pomeron
models and being given by the shape from the inclusive case in the soft color 
interaction model.
Thus, we reach the conclusion that the study of the double diffractive 
electroweak boson production   
at the LHC is an ideal test of the two distinct QCD mechanisms responsible 
for diffractive processes
in hadronic collisions.

\section{$W$ boson rapidity asymmetry}

\subsection{Non-diffractive case}

\begin{figure}[t]
\begin{center}
\includegraphics[width=6.5cm]{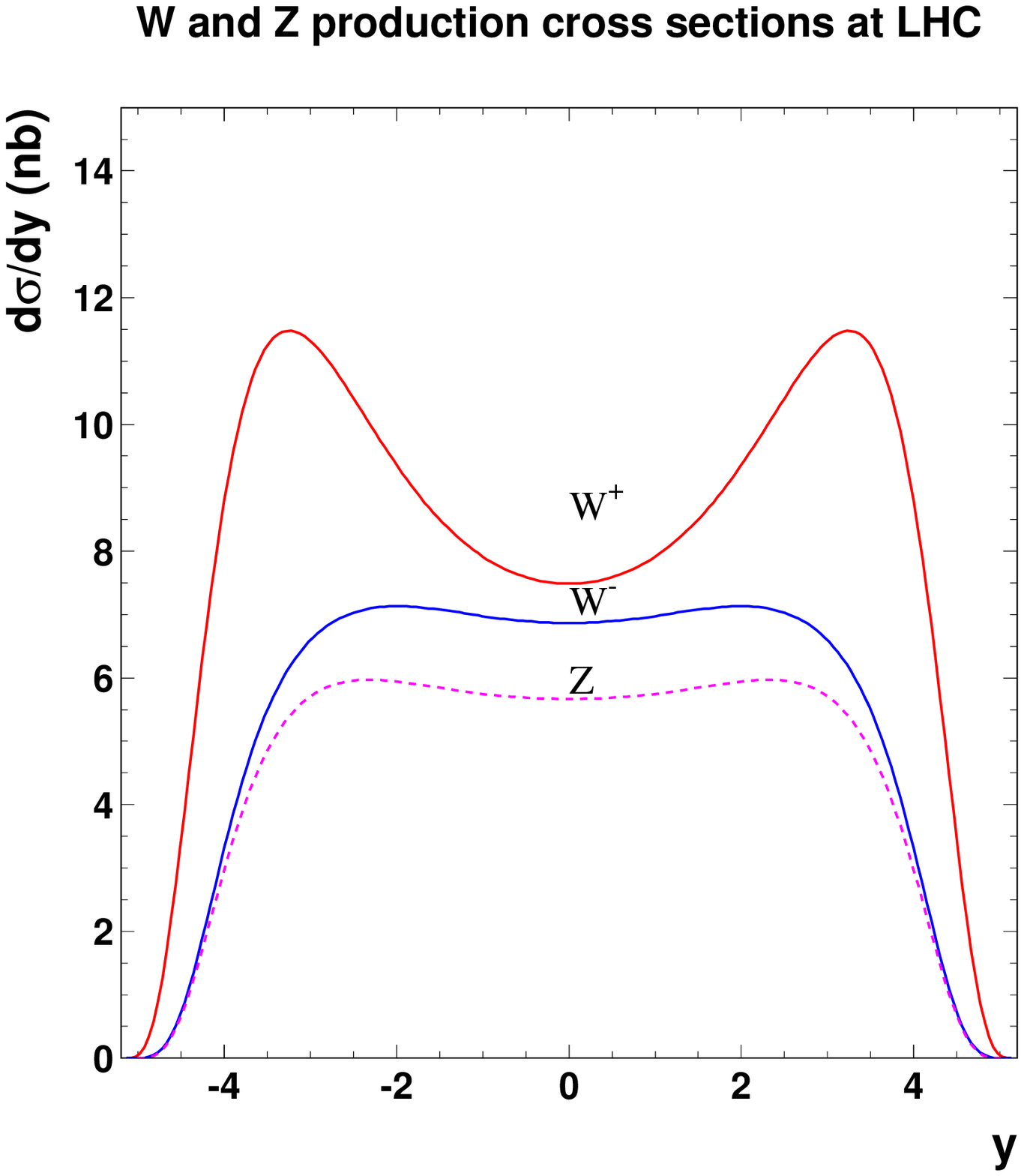}
\includegraphics[width=6.5cm]{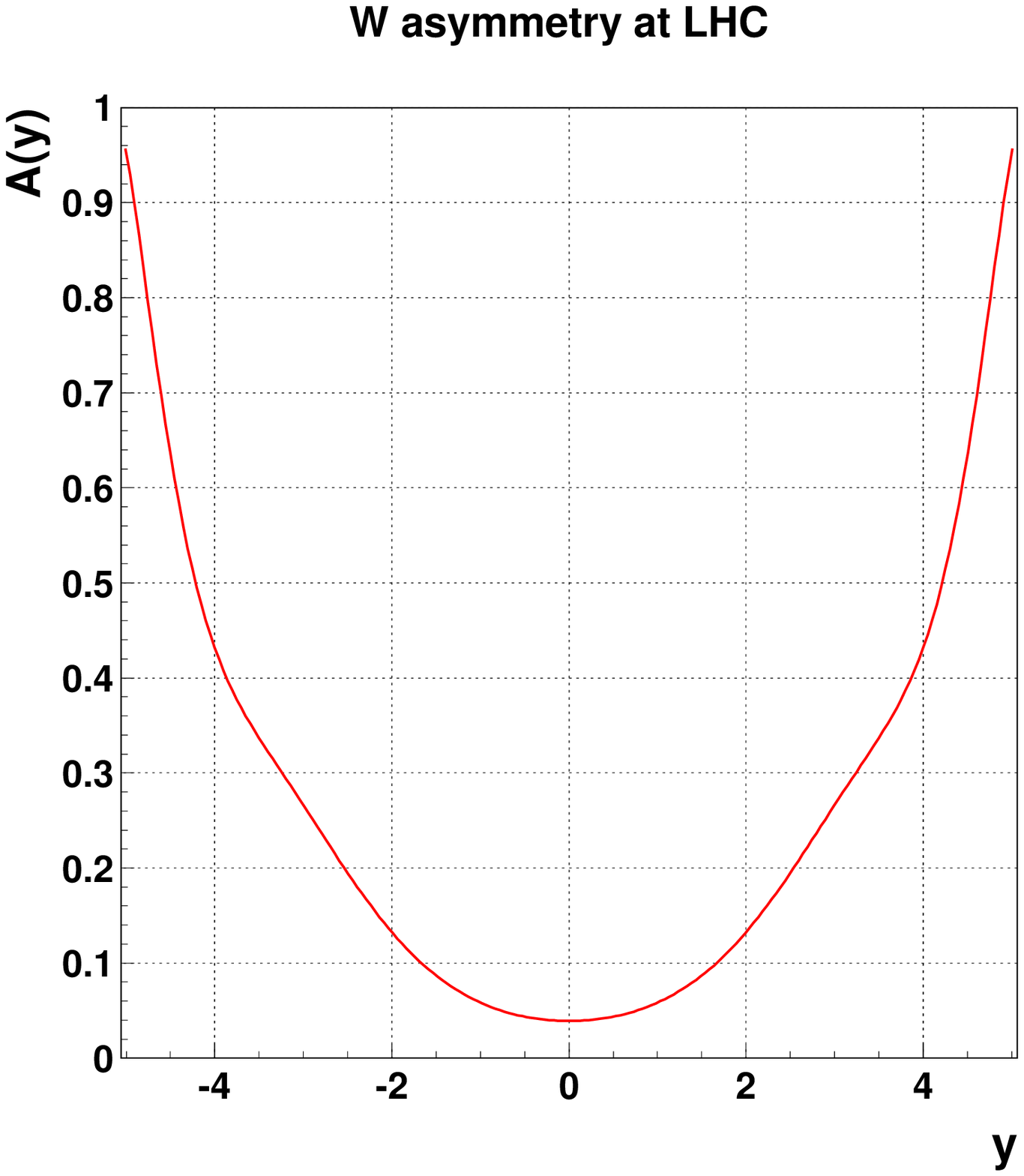}
\caption{The $W$ and $Z$ boson production cross sections at the LHC as 
a  function of the boson rapidity $y$ (left) and the corresponding 
$W$ boson rapidity asymmetry (right) for the  LO MSTW08 parton distributions.}
\label{fig:1}
\end{center}
\end{figure}

The electroweak vector bosons are produced in the $pp$ scattering from 
annihilation of two quarks. The  boson rapidity $y$ is determined by the 
longitudinal proton momentum fractions, $x_1$ and $x_2$, carried by the 
colliding quarks
\begin{eqnarray}
x_1=\frac{M_W}{\sqrt{s}} {\rm e}^{y}\,,~~~~~~~~~~~~~~~~~~~~~
x_2=\frac{M_W}{\sqrt{s}} {\rm e}^{-y}
\label{kin}
\end{eqnarray}
where $M_W$ is  the $W$ boson mass and $\sqrt{s}$ is invariant energy of 
scattering protons.
From the conditions $0\le x_{1,2}\le 1$, we find the following bound for the $W$ rapidity
\be
-y_{min}\le y \le y_{min}
\ee
where $y_{min}=\ln(\sqrt{s}/M_W)$. 
The  measured inclusively $W^{\pm}$ rapidity asymmetry,
\begin{equation}
A_{\rm incl}(y)=\frac{d\sigma_{W^+}(y)/dy-d\sigma_{W^-}(y)/dy}{d\sigma_{W^+}(y)/dy+d\sigma_{W^-}(y)/dy}\,,
\end{equation}
is determined mainly by the $u_p$ and $d_p$ quark distributions in the colliding protons. 
With a simplifying assumption
that the anti-quark distributions in the proton are equal, $\ubar=\dbar$, 
we have  \cite{Qcdandcollider}
\be\label{eq:aincl}
A_{\rm incl}(y)=\frac{(u_p(x_1)-d_p(x_1))\,\ubar_p(x_2)\,+\, \ubar_p(x_1)\,(u_p(x_2)-d_p(x_2))}
{(u_p(x_1)+d_p(x_1))\,\ubar_p(x_2)\,+\, \ubar_p(x_1)\,(u_p(x_2)+d_p(x_2))}
\ee
where
the quark distributions also depend  on a hard scale given by the $W$ boson mass $\mu=M_W$.
The measured non-zero value of $A_{\rm incl}(y)$ is explained by different distributions 
of $u$ and $d$ quarks in the proton:
$u_p\ne d_p$.
Thus, the $W$ rapidity asymmetry is a good observable to pin down these distributions and is usually included
in global fits of parton distribution functions in a nucleon. The $W$ and $Z$ boson production cross sections
and the corresponding $W$ rapidity asymmetry are shown in Fig.~\ref{fig:1}
in the non-diffractive case for the proton (with the strange quark contribution included), where the leading order MSTW08 
parametrization of PFDs was used \cite{Martin:2009iq}.

\subsection{Diffractive case}

In the double diffractive production, the electroweak bosons
are parts of a diffractive system with mass $M_D$ which is separated in rapidity from the scattered
protons by two gaps of the size $\ln(1/\xi_1)$ and $\ln(1/\xi_2)$, where $\xi_{1,2}$ are small fractions of
the incoming proton momenta transferred into the diffractive system. They obey the following relation
\be
\xi_1\xi_2=\frac{M_D^2}{s}\,,~~~~~~~~~~~~~\xi_{1,2}\ll 1\,.
\ee  
Thus, rapidity of the diffractively produced $W$ boson stays  in the range
\be
-y_{min}+\ln(1/\xi_1)\,\le\, y \,\le\, y_{min}-\ln(1/\xi_2)\,.
\ee
The relevant $W$ rapidity asymmetry is defined now as the ratio of triple differential cross sections 
\be\label{eq:adiff}
A_D(y,\xi_1,\xi_2)=\left(\frac{d\sigma_{W^+}}{dyd\xi_1d\xi_2}-\frac{d\sigma_{W^-}}{dyd\xi_1d\xi_2}\right)
\Big/\left(\frac{d\sigma_{W^+}}{dyd\xi_1d\xi_2}+\frac{d\sigma_{W^-}}{dyd\xi_1d\xi_2}\right).
\ee

\section{QCD interpretations of diffractive $W$ boson rapidity asymmetry}
\label{rpm}

The electroweak boson mass  is a hard scale which allows 
for perturbative QCD interpretation of the $W$ or $Z$ production. 
However, in the diffractive production of electroweak vector bosons 
there are several approaches to the nature of the vacuum quantum number exchange 
which leads to rapidity gaps. 

In the resolved Pomeron exchange interpretation, 
the Pomeron is endowed with a partonic structure described 
by the Pomeron parton distributions.
The  double diffractive  processes can then be qualified as a double Pomeron exchange (DPE) with
vacuum quantum numbers. In this case,  the ordinary parton distributions in the proton are replaced by diffractive parton distribution functions (DPDFs), $q_D(x,\xi)$ for quarks and $g_D(x,\xi)$ for gluons. They  encode the information about  momentum fractions $\xi$ transferred from  the initial proton into the diffractive system
\cite{RPE,Berera:1994xh,Berera:1995fj,Trentadue:1993ka,Collins:1994zv}. 
An additional assumption is usually made that the diffractive parton distributions have a factorized form
\be\label{eq:fact}
q_D(x,\xi,\mu)=\Phi(\xi)\,q_\funp(\beta,\mu)\,,
\ee
where $\beta=x/\xi$ is a fraction of the Pomeron momentum carried by a quark participating 
in the diffractive scattering. We also indicated the hard scale dependence of these distributions, $\mu=M_W$ in our case.
The quantity $q_\funp(\beta,\mu)$ is called a Pomeron quark distribution, and $\Phi(\xi)$ represents a Pomeron flux.
In this way, the vacuum exchange aspect of diffractive processes separated from the hard scattering of two quarks which produces the $W$ boson. However, the assumption about the factorization (\ref{eq:fact}) is not necessary in our considerations.

\begin{figure}[t]
\begin{center}
\includegraphics[width=12cm]{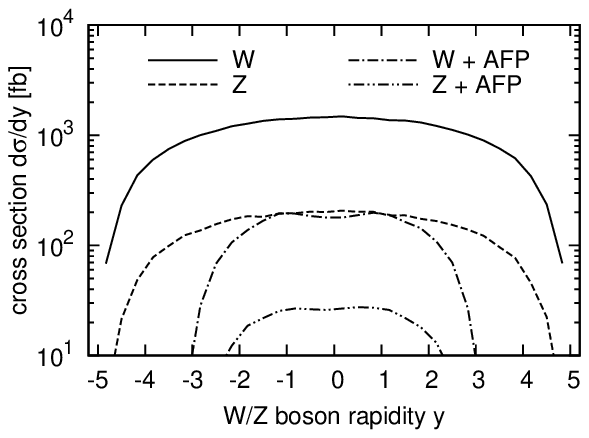}
\includegraphics[width=12cm]{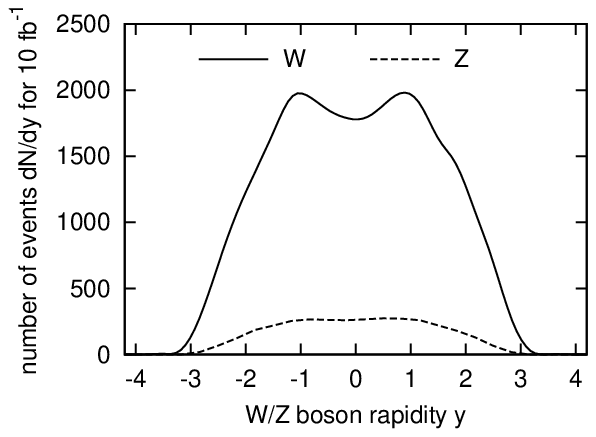}
\caption{Top: diffractive $W$ and $Z$ boson production cross sections as a function
of  rapidity in the DPE
model with and without taking into account the AFP acceptance (see the text). Bottom: the same
cross sections in the SCI model.}
\label{fig:2}
\end{center}
\end{figure}

The leading order cross sections for the DPE models of the $W$ boson production are given by the  inclusive case
formula \cite{Qcdandcollider}  with the diffractive quark distributions
\beeq
\frac{d^3\sigma_{W^+}}
{dyd\zeta_1d\zeta_2}\eq \sigma_0^W\,V_{ud}^2\left\{u_D(x_1,\xi_1)\,\dbar_D(x_2,\xi_2)\,+\,
\dbar_D(x_1,\xi_1)\,u_D(x_2,\xi_2)  \right\} S^2   
\label{eq1}
\\
\frac{d^3\sigma_{W^-}}
{dyd\zeta_1d\zeta_2}\eq \sigma_0^W\,V_{ud}^2\left\{ d_D(x_1,\xi_1)\,\ubar_D(x_2,\xi_2)\,+\,
\ubar_D(x_1,\xi_1)\,d_D(x_2,\xi_2)      \right\} S^2  
\label{eq2}
\eeeq
where 
$\sigma_0^W={2\pi G_F M_W^2}/{(3\sqrt{2}s)}$ and $V_{ud}$ is 
the CKM matrix element. In addition, we multiply the cross sections by a survival probability \cite{Bjorken:1992er}
of the DPE process, 
$S^2$, which might be a complicated function of the fractions $\xi_{1,2}$.
We also neglected the Cabbibo suppressed $s$ quark part of the $W$ 
boson production cross sections \cite{Qcdandcollider} . 
Note that adding NLO and NNLO corrections to cross section expressions above \cite{Catani:2009sm} 
is not important for the idea developed in this paper on asymmetries. 

The $W$ boson asymmetry (\ref{eq:adiff}) computed from the above cross sections is given by  
formula (\ref{eq:aincl}) with the  ordinary quark distributions replaced by the diffractive ones.
Since the Pomeron carries vacuum quantum numbers, 
it is expected to be made of gluons and sea quarks.
Then, diffractive PDFs  in eqs.~(\ref{eq1}) and (\ref{eq2}) have to follow the relations:
$u_D=\ubar_D$ and $d_D=\dbar_D$. Thus, 
the $W^{\pm}$ boson production asymmetry in the DPE models equals zero,
\begin{equation}
A_D(y,\xi_1,\xi_2)=0\,,
\end{equation}
independent of the rapidity gap sizes determined by the fractions $\xi_{1,2}$.
 
The $W^{\pm}$ and $Z$ cross sections are displayed in 
Fig.~\ref{fig:2} 
(top) for the pomeron-based models with and without taking into account the
ATLAS Forward Proton (AFP) detectors \cite{fpmc,afp}. The AFP project consists in installing
forward proton detectors located at 220 m and 420 m from the ATLAS interaction
point. In the first phase of the project, only the detectors at 220 m are
considered which leads to an acceptance in $\xi$, the proton momentum loss, of $0.01<\xi<0.15$ and this
acceptance is assumed in the following of the paper. Let us note that the
acceptance increases down to 0.002 if detectors at 420 m are added, which is
fundamental to detect the exclusive production of lower mass objects, such as 
the Higgs bosons \cite{higgs,kepka}. In Fig.~\ref{fig:2} is displayed the number of
$Z$ and $W$ events observed in the AFP acceptance for a luminosity of 10
fb$^{-1}$, which is quite low at the LHC \cite{afp}. We note the high number of events
measured even at a relatively low luminosity which allows to probe quantitatively
the different models of diffraction, as we will see in the following. 
\begin{figure}[t]

\begin{center}
\includegraphics[width=9.5cm]{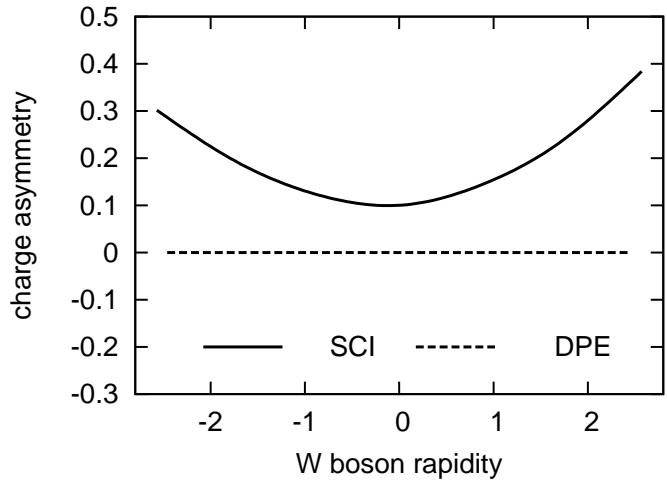}
\caption{$W$ asymmetries as a function of the $W$ boson rapidity $y$ 
in the pomeron based  (DPE) and soft color interaction (SCI) models, using  the AFP
acceptance. We note the flat distribution for the DPE models and a strong
$y$ dependence for the SCI models.} 
\label{fig:2b}
\end{center}
\end{figure}

The soft color interaction approach (SCI) modifies the color reconnection at the hadronization level
in order to produce rapidity gaps. Then, the diffractive PDFs are not needed and 
up to correction factors originating from features of the hadronization process,
the $W$ boson production cross sections read as for the standard inclusive case
\beeq
\frac{d^3\sigma_{W^+}}{dy\,d\zeta_1d\zeta_2} &\propto& 
\sigma_0^W\,V_{ud}^2\left\{u_p(x_1)\,\dbar_p(x_2)\,+\,\dbar_p(x_1)\,u_p(x_2)  \right\}
\\
\frac{d^3\sigma_{W^-}}{dy\,d\zeta_1d\zeta_2} &\propto& 
\sigma_0^W\,V_{ud}^2\left\{ d_p(x_1)\,\ubar_p(x_2)\,+\,\ubar_p(x_1)\,d_p(x_2)      \right\}
\eeeq
with the same proportionality coefficient. Thus, we expect the asymmetry (\ref{eq:adiff}) to be equal or be very close to the non-zero value given by the standard $W$ rapidity  asymmetry (\ref{eq:aincl}),
\be
A_D(y,\xi_1,\xi_2)=A_{\rm incl}(y)\,,
\ee
independent of the rapidity gap sizes. The $W^{\pm}$  cross sections in the SCI approach are shown  
in Fig.~\ref{fig:2} (bottom). They have have to be multiplied by a rapidity gap survival factor \cite{Bjorken:1992er}.
Note that the global factors in the SCI model that multiply the above expressions
disappear in the asymmetry ratio. In this way, by measuring the $W$ rapidity asymmetry in the 
double diffractive processes, we are able to discriminate between the DPE and SCI mechanisms of diffraction in the $pp$
scattering.

We summarize this conclusion in Fig.~\ref{fig:2b} where
the $W$ asymmetries are shown  in the case of the pomeron 
and the soft color interaction based models. The $y$-dependence of the $W$ asymmetry is expected to be flat and close to zero in the DPE models  whereas for the SCI models it is close to the non-diffractive $W$ asymmetry
in the $pp$ scattering case. Distinguishing between both models will be easy at the LHC, given the
cross section of those processes and the luminosity available.

\section{$W/Z$ production ratio: distinction between DPE and SCI models}

We also consider the $Z$ production rate, determined in the inclusive case by the leading order cross section
\be\label{eq:zcs}
\frac{d\sigma_{Z}}{dy}= \sigma_0^Z\left\{C_u u_p(x_1^\prime)\,\ubar_p(x_2^\prime)
\,+\,C_d \,d_p(x_1^\prime)\,\dbar_p(x_2^\prime) \,+\,(x_1^\prime\leftrightarrow x_2^\prime)\right\}
\ee
with $\sigma_0^Z={2\pi G_F M_Z^2}/{(3\sqrt{2}s)}$, $C_{u,d}=V^2_{u,d}+A^2_{u,d}$ where  
$V_{u,d}= T_{u,d}^3-2Q_{u,d}\sin^2\theta_W$ and $A_{u,d}= T_{u,d}^3$ are the vector and axial couplings of  $u$ and $d$ quarks to   $Z$ boson. Notice  that the PDFs are computed at the factorization scale $\mu=M_Z$, and the momentum fractions are also related to $Z$ boson mass, $x_1^\prime={M_Z}{\rm e}^{\pm y}/{\sqrt{s}}$. 
The $W$ to $Z$ production ratio,
\be
R_{inc}(y)\equiv\frac{1}{2}\left(\frac{d\sigma_W^+}{dy}+\frac{d\sigma_W^+}{dy}\right)\Big/\frac{d\sigma_Z}{dy}\,,
\ee
in the inclusive case reads
\be\label{eq:rincl}
R_{\rm incl}(y)\,=\,
\frac{M_W^2 V_{ud}^2}{M_Z^2}
\frac{u_p(x_1)\,\dbar_p(x_2)+\dbar_p(x_1)\,u_p(x_2)+d_p(x_1)\,\ubar_p(x_2)+\ubar_p(x_1)\,d_p(x_2)}
{C_u u_p(x_1^\prime)\,\ubar_p(x_2^\prime)+C_d \,d_p(x_1^\prime)\,\dbar_p(x_2^\prime)
 +(x_1^\prime\leftrightarrow x_2^\prime)}.
\ee
The distributions in the numerator are taken at the scale 
$\mu=M_W$ (and $x_{1,2}={M_W}{\rm e}^{\pm y}/{\sqrt{s}}$) while for those in the denominator
$\mu=M_Z$. 

\begin{figure}[t]
\begin{center}
\includegraphics[width=8.5cm]{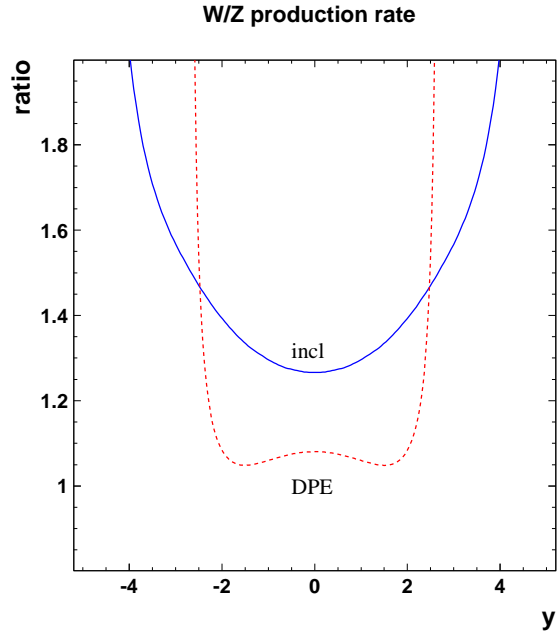}
\caption{The $W$ to $Z$ production  ratio in the DPE and SCI models (see text).}
\label{fig:3}
\end{center}
\end{figure}

In the DPE model, in  eq.~(\ref{eq:zcs}) the ordinary proton PDFs are replaced by the diffractive PDFs and additionally, the cross section is multiplied by a survival factor $S^2$,
\be
\frac{d^3\sigma_{Z}}{dy\,d\xi_1d\xi_2} =\sigma_0^Z
\left\{C_u u_D(x_1^\prime,\xi_1)\,\ubar_D(x_2^\prime,\xi_2)\,+\,C_d \,d_D(x_1^\prime,\xi_1)\,\dbar_D(x_2^\prime,\xi_2) \,+\,
((x_1^\prime,\xi_1)\leftrightarrow (x_2^\prime,\xi_2))\right\} S^2\,.
\ee
The corresponding $W$ to $Z$ production ratio, defined by the formula
\be
R_{D}(y,\xi_2,\xi_2)\equiv
\frac{1}{2}\left(\frac{d^3\sigma_W^+}{dy\,d\xi_1d\xi_2}+\frac{d^3\sigma_W^-}{dy\,d\xi_1d\xi_2}\right)
\Big/\frac{d^3\sigma_Z}{dy\,d\xi_1d\xi_2}\,,
\ee
has the form (\ref{eq:rincl}) with the diffractive PDFs. The multiplicative survival factor $S^2$ cancels out
in this ratio.  The ratio $R_{D}$ for the DPE model is shown as the dashed line in Fig.~\ref{fig:3}.
Assuming the flavour symmetry of the diffractive PDFs, $u_D=\ubar_D=d_D=\dbar_D$, we find an almost constant 
ratio in the central rapidity region 
\be\label{eq:rdpe}
R_{D}(y,\xi_2,\xi_2)
\,\approx\, \frac{M_W^2 V_{ud}^2}{M_Z^2}\frac{1}{C_u+C_d}\,.
\ee
A much  larger deviation from a constant value occurs close to the rapidity gap edges, see Fig.~\ref{fig:3}.

In the SCI model, however, the ratio depends on rapidity through the ordinary quark distributions in the proton,
and we obtain
\be\label{eq:rsci}
R_{D}(y,\xi_2,\xi_2)=R_{\rm incl}(y)\,.
\ee
This ratio is shown in Fig.~\ref{fig:3} as the solid line.
We notice again that the $W$ to $Z$ cross section ratio  is an additional observable
allowing to distinguish between the  DPE and SCI models.


\section{Test of  flavor symmetry of diffractive PDFs}

The almost constant ratio $R_{D}$ in the DPE models was obtained under the assumption of full
flavor symmetry of the DPDFs. Let us note that this hypothesis can not be tested in standard QCD fits of diffractive PDFs or in any of the diffractive processes examined with HERA data \cite{hera,hera2}.
Thus, if the measurements at the LHC show the results different from
our conclusions, eqs.~(\ref{eq:rdpe}) or (\ref{eq:rsci}), there is an interesting possibility that the Pomeron
parton distribution, see eq.~(\ref{eq:fact}), are not fully flavor symmetric, e.g. 
\be
u_\funp(\beta)=\ubar_\funp(\beta)\,\ne\,d_\funp(\beta)=\dbar_\funp(\beta)\,.
\ee

In such a case the diffractive ratio takes the following form
\be\label{eq:rpom}
R_{diff}(y,\xi_2,\xi_2)=\frac{M_W^2V_{ud}^2}{M_Z^2}
\frac{u_\funp(\beta_1,M_W)\,d_\funp(\beta_2,M_W)+d_\funp(\beta_1,M_W)\,u_\funp(\beta_2,M_W)}
{C_u u_\funp(\beta_1^\prime,M_Z)\,u_\funp(\beta_2^\prime,M_Z)
+C_d \,d_\funp(\beta_1^\prime,M_Z)\,d_\funp(\beta_2^\prime,M_Z)}
\ee
where   $\beta_{1,2}=x_{1,2}/\xi_{1,2}$ and  $\beta_{1,2}^\prime=x_{1,2}^\prime/\xi_{1,2}$ are the Pomeron momentum fractions carried by the quarks producing the electroweak bosons.
The above expression  provides a direct sensitivity of the the $d_\pom/u_\pom$ ratio, albeit for the quark distributions taken at two different scales, $\mu=M_{W,Z}$ .
In case of flavor symmetry, the ratio is 
expected to be approximately a constant, if not, a non--trivial shape may be obtained.

The measurement of the $W$ to $Z$ cross section ratio is sensitive to the $u$,
$d$, $s$ quark densities in the Pomeron and especially to their ratios. The H1 and ZEUS experiments
measured the structure of the pomeron \cite{hera,hera2}. The fits always
assume $u_\funp=d_\funp=s_\funp$ since data are not sensitive to their difference. The measurement
of the $W$ to $Z$ cross section will allow to probe this assumption. 

\begin{figure}[t]
\begin{center}
\includegraphics[width=10cm]{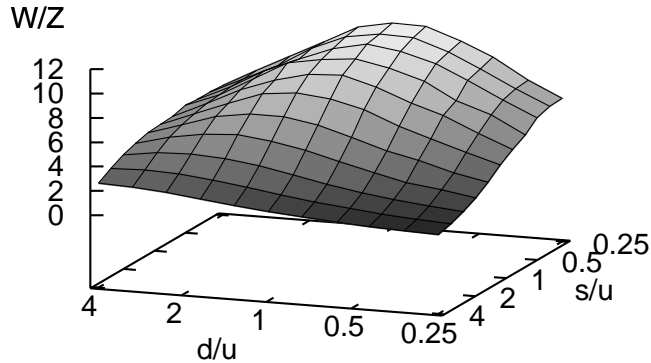}
\caption{The effect of varying the $d/u$ and $s/u$ quark density ratio in the Pomeron on the $W/Z$
cross section ratio keeping $u+d+s$ constant.
}
\label{fig4}
\end{center}
\end{figure}

\begin{figure}[htbp]
\begin{center}
\includegraphics[width=8cm]{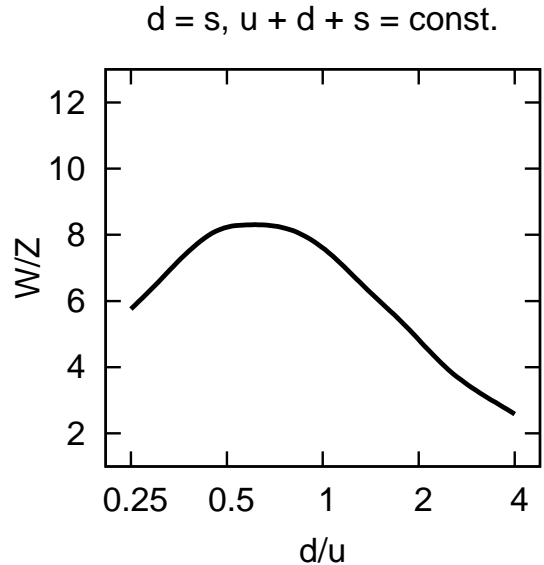}
\caption{The effect of varying the $d/u$ quark density 
ratio in the Pomeron on the $W/Z$
cross section ratio keeping $u+d+s$ constant and assuming $d=s$. }
\label{fig4bb}
\end{center}
\end{figure}

In
Fig~\ref{fig4}, we display the $W$ to $Z$ cross section ratio as a function of
$s_\funp/u_\funp$ and $d_\funp/u_\funp$ ratios while keeping the sum $u_\funp+d_\funp+s_\funp$ constant. We note the strong
dependence of the $W$ to $Z$ cross section ratio on the quark density ratio,
which will allow to probe the assumption $u_\funp=d_\funp=s_\funp$ using LHC data.
In order to show more precisely this dependence, we show in Fig.~\ref{fig4bb} 
one projection along a vertical axe:
we display the cross section ratio varying for instance 
$d_\funp/u_\funp$ assuming $d_\funp=s_\funp$ and $u_\funp+d_\funp+s_\funp$ constant as usual (since this is well
constrained by the QCD fits performed at HERA). We notice that the
effect of the cross section ratio can be more than a factor of four while
varying the quark densities, which shows the potential of such a measurement at
the LHC.

\section{Conclusion}

We have shown that the  double diffractive electroweak vector boson production in the $pp$ collisions  
at the LHC is an ideal probe of QCD based mechanisms of diffraction. 
Assuming the resolved Pomeron  model, the $W$ production
asymmetry in rapidity have been shown to be exactly zero for all rapidities. 
In other approaches, like  the soft color interaction model, in which  soft gluon exchanges are responsible for diffraction,  the asymmetry is non-zero and equal to that in
the inclusive $W$ production. In the same way, the ratio of the $W$ to $Z$ boson production is independent of rapidity in the models with resolved Pomeron and flavor symmetry in contrast to the predictions of the soft color interaction model.
The sensitivity to the ratio of the $d$ to $u$ quarks in the Pomeron, using the $W$ to $Z$ cross sections ratio, has been studied. Large variations have been found.

\begin{acknowledgments}
This work was supported the grant of Polish National Science Center no.
DEC-2011/01/B/ST2/03915.

\end{acknowledgments}



\begin{thebibliography}{99}

\bibitem{Abe:1994rj}
CDF, F.~Abe {\em et~al.},
Phys. Rev. Lett. {\bf 74}, 850 (1995), [hep-ex/9501008].

\bibitem{Abazov:2008qv}
D0, V.~M. Abazov {\em et~al.},
Phys. Rev. Lett. {\bf 101}, 211801 (2008), [0807.3367].

\bibitem{Aaltonen:2009ta}
CDF, T.~Aaltonen {\em et~al.},
Phys. Rev. Lett. {\bf 102}, 181801 (2009), [0901.2169].


\bibitem{atlas}
  G.~Aad {\it et al.}  [ATLAS Collaboration],
  Phys.\ Lett.\  B {\bf 701} (2011) 31
  [arXiv:1103.2929 [hep-ex]].

\bibitem{cms}
  S.~Chatrchyan {\it et al.} [ CMS Collaboration ],
  JHEP {\bf 1104 } (2011)  050.
  [arXiv:1103.3470 [hep-ex]].

\bibitem{Catani:2009sm}
  S.~Catani, L.~Cieri, G.~Ferrera, D.~de Florian, M.~Grazzini,
  Phys.\ Rev.\ Lett.\  {\bf 103 } (2009)  082001.
  [arXiv:0903.2120 [hep-ph]].

\bibitem{GolecBiernat:2009pj}
  K.~Golec-Biernat, A.~Luszczak,
  Phys.\ Rev.\  {\bf D81 } (2010)  014009.
  [arXiv:0911.2789 [hep-ph]].


\bibitem{general}
M. Boonekamp, F. Chevallier, C. Royon, L. Schoeffel. Feb 2009. 92 pp.
Published in Acta Phys.Polon. B40 (2009) 2239-2321


\bibitem{kupco}
Alexander Kupco, Christophe Royon, Robert B. Peschanski,
Phys.Lett. B {\bf 606} (2005) 139.


\bibitem{RPE}
J.~C.~Collins,
Phys.\ Rev.\  D {\bf 57} (1998) 3051
[Erratum-ibid.\  D {\bf 61} (2000) 019902];
G.~Ingelman and P.~E.~Schlein,
Phys.\ Lett.\  B {\bf 152} (1985) 256.


\bibitem{SCI}
A. Edin, G. Ingelman, J. Rathsman, Phys. Lett. {\bf B366} (1996) 371.

\bibitem{Qcdandcollider}
R.~K. Ellis, W.~J. Stirling and W.~B. R.,
\newblock QCD and Collider Physics, {\it Cambridge University Press}  (1996).


\bibitem{Martin:2009iq}
  R.~S.~Thorne, A.~D.~Martin, W.~J.~Stirling and G.~Watt,
  arXiv:0907.2387 [hep-ph].

\bibitem{Berera:1994xh}
A.~Berera and D.~E. Soper,
Phys. Rev. {\bf D50}, 4328 (1994), [hep-ph/9403276].

\bibitem{Berera:1995fj}
A.~Berera and D.~E. Soper,
Phys. Rev. {\bf D53}, 6162 (1996), [hep-ph/9509239].

\bibitem{Trentadue:1993ka}
L.~Trentadue and G.~Veneziano,
Phys. Lett. {\bf B323}, 201 (1994).

\bibitem{Collins:1994zv}
J.~C. Collins, J.~Huston, J.~Pumplin, H.~Weerts and J.~J. Whitmore,
Phys. Rev. {\bf D51}, 3182 (1995), [hep-ph/9406255].

\bibitem{Bjorken:1992er}
J.~D. Bjorken,
Phys. Rev. {\bf D47}, 101 (1993).


\bibitem{fpmc}
M. Boonekamp, V. Jur\'{a}nek, O. Kepka, C. Royon, Proceedings of the
Workshop of the Implications of HERA for LHC
physics, DESY-PROC-2009-02;
  arXiv:0903.3861;
M. Boonekamp, A. Dechambre, V. Juranek, O. Kepka, M. Rangel, C. Royon, R. Staszewski, arXiv:1102.2531.


\bibitem{afp} M. G. Albrow et al., JINST 4 (2009) T10001; C. Royon,
Proceedings of the DIS 2007 workshop, Munich, preprint arXiv:0706.1796.

\bibitem{higgs} V.A. Khoze, A.D. Martin, M.G. Ryskin, Eur. Phys. J. {\bf
C19} (2
001) 477;
Eur. Phys. J. C{\bf 23} (2002) 311;
Eur. Phys. J. C{\bf 24} (2002) 581; M. Boonekamp, R. Peschanski, C. Royon,
Phys.
  Rev. Lett. {\bf  87 }
(2001)
251806; Nucl. Phys. B{\bf 669} (2003) 277;
A. Dechambre, O. Kepka, C. Royon, R. Staszewski,
Phys. Rev. D {\bf 83} 054013 (2011).

\bibitem{kepka}
O.~Kepka and C.~Royon,
forward
Phys.\ Rev.\  D {\bf 78} (2008) 073005
[arXiv:0808.0322 [hep-ph]].
E. Chapon, O. Kepka, C. Royon, Phys. Rev. {\bf D81} (2010) 074003;
J. de. Favereau et al., preprint arXiv:0908.2020.


\bibitem{hera}
A.~Aktas {\it et al.}  [H1 Collaboration],
Eur.\ Phys.\ J.\  C {\bf 48} (2006) 715;
S.~Chekanov  [ZEUS Collaboration],
Nucl.\ Phys.\  B {\bf 800} (2008) 1.

\bibitem{hera2}
C.~Royon, L.~Schoeffel, S.~Sapeta, R.~B.~Peschanski and E.~Sauvan,
Nucl.\ Phys.\  B {\bf 781} (2007) 1.




\end{thebibliography}
\end{document}